\documentclass{aastex631}

\newcommand\kms{km\,$\rm s^{-1}$}

\submitjournal{ApJ}

\shorttitle{Tracing the Origin of Moving Groups}
\shortauthors{Yang et al.}
\graphicspath{{./}{figures/}}

\begin{document}

\title{Tracing the Origin of Moving Groups. III. Detecting Moving Groups in The LAMOST DR7}

\correspondingauthor{Jingkun Zhao}
\email{zjk@bao.ac.cn}

\author[0000-0001-7609-1947]{Yong Yang}
\affiliation{Key Laboratory of Optical Astronomy, National Astronomical Observatories, Chinese Academy of Sciences, Beijing 100101, China}
\affiliation{School of Astronomy and Space Science, University of Chinese Academy of Sciences, Beijing 100049, China},
\author[0000-0003-2868-8276]{Jingkun Zhao}
\affiliation{Key Laboratory of Optical Astronomy, National Astronomical Observatories, Chinese Academy of Sciences, Beijing 100101, China},
\author[0000-0003-1352-7226]{Jiajun Zhang}
\affiliation{Key Laboratory of Optical Astronomy, National Astronomical Observatories, Chinese Academy of Sciences, Beijing 100101, China}
\affiliation{School of Astronomy and Space Science, University of Chinese Academy of Sciences, Beijing 100049, China},
\author{Xianhao Ye}
\affiliation{Key Laboratory of Optical Astronomy, National Astronomical Observatories, Chinese Academy of Sciences, Beijing 100101, China}
\affiliation{School of Astronomy and Space Science, University of Chinese Academy of Sciences, Beijing 100049, China},
\author{Gang Zhao}
\affiliation{Key Laboratory of Optical Astronomy, National Astronomical Observatories, Chinese Academy of Sciences, Beijing 100101, China}
\affiliation{School of Astronomy and Space Science, University of Chinese Academy of Sciences, Beijing 100049, China}



\begin{abstract}

We revisit the moving groups (MGs) in the solar neighborhood with a sample of 91969 nearby stars constructed from LAMOST DR7. Using the wavelet technique and Monte Carlo simulations, five MGs together with a new candidate located at $V \simeq$ -130 \kms\ are detected simultaneously in $V-\sqrt{U^2+2V^2}$ space. Taking into account the other known MGs, we conclude that MGs in the Galactic disk are spaced by approximately 15 $\sim$ 25 \kms\ along $V$ velocity. The origin of detected MGs is analysed through the distributions of [Fe/H]$-$[Mg/Fe] and ages. Our results support attributing the origin to the continuous resonant mechanisms probably induced by the bar or spiral arms of the Milky Way.

\end{abstract}



\section{Introduction} \label{introduction}

Moving groups (MGs) are detected as kinematic substructures in the solar neighborhood whose member stars share similar velocity components. Different kinds of hypotheses have been put forward to interpret the origin of them. It was early believed that they came from dissolving open clusters \citep[e.g., HR1614;][]{2007AJ....133..694D}. However, this interpretation is incompatible with most of known MGs for the inhomogeneity of age and chemistry within them \citep[e.g.,][]{2005A26A...430..165F,2012MNRAS.425.3188R,2019A26A...631A..47K}. Later, dynamical mechanism of internal resonances caused by the Galactic bar or spiral arms was proposed \citep[e.g.,][]{2000AJ....119..800D,2005AJ....130..576Q,2019A26A...626A..41M}. Specifically, the existence of Hercules is highly consistent with the effects of the bar resonances \citep{2007ApJ...655L..89B}. Some structures with low angular momenta below Hercules were reproduced, considering spiral arms plus a bar with a pattern speed of $\Omega _b$ = 45 \kms\,$\rm kpc^{-1}$ and a hot disk with a velocity dispersion of $\sim$ 40 \kms\ at $R_{\sun}$ = 8.5 kpc as the initial condition \citep[Figure 1(i) in][]{2009ApJ...700L..78A}. Some retrograde MGs pertaining to the Galactic halo were also related to resonant orbits created by the bar \citep{2019IAUS..344..134S}. Furthermore, MGs can be explained as relics of disrupted satellite galaxies or products of perturbations by external accretion events \citep{2009MNRAS.396L..56M}. For example, \citet{2004ApJ...601L..43N} attributed the Arcturus group to remnants of a dwarf galaxy merged with the Milky Way since a tight sequence in the [Fe/H]$-$[$\alpha$/Fe] plane was found, although this argument has been disproved using more unbiased data \citep{2019A26A...631A..47K}.

Nowadays, the Gaia mission \citep{2018A26A...616A...1G, 2021A26A...649A...1G}, along with spectroscopic surveys such as LAMOST \citep{2012RAA....12.1197C,2006ChJAA...6..265Z,2012RAA....12..723Z,2021SCPMA..6439562Z} and APOGEE \citep{2017AJ....154...94M,2020ApJS..249....3A}, has provided ample stellar astrometric, photometric and spectroscopic information, and unprecedented details of MGs have been revealed. These large data sets also allow the derivation of stellar age, which plays another crucial role in order to better understand the MGs. In \citet{2008A26A...490..135A}, ages were used to study the evolutionary state of the nearby kinematic substructures. The origin of Hercules was investigated through the age distributions in \citet{2007ApJ...655L..89B, 2014A26A...562A..71B}. 

In \citet[paper I]{2018ApJ...863....4L} and \citet[paper II]{2018ApJ...868..105Z}, the origins of the $\gamma$ Leo moving group and LAMOST-N1 \citep{2015RAA....15.1378Z} were investigated through detailed abundance analysis. In this work, we detect MGs in the solar neighborhood with samples constructed from LAMOST DR7. We further discuss the origin of the MGs by analysing their chemical properties and ages. Section \ref{data} describes the data process. Section \ref{detection} characterizes MGs detection in detail. Section \ref{chem_age} analyses the origin of MGs through chemistry and age. A summary is presented in Section \ref{summary}.

\section{Data} \label{data}

The LAMOST DR7 provides stellar atmospheric parameters and radial velocities for 6,159,427 stars, including 94,908 A type stars, 1,893,014 F type stars, 3,099,821 G type stars and 1,071,684 K type stars. The data are cross-matched with Gaia eDR3 to get proper motion. Distance comes from Bayesian estimate \citep{2021AJ....161..147B}, which is derived using the parallax, colour and apparent magnitude of a star (called ``photogeometric distance").

Heliocentric velocities and corresponding uncertainties, together with angular momenta and actions, are computed using $galpy$\footnote{Available at \url{http://github.com/jobovy/galpy}} Python package \citep{2015ApJS..216...29B} by adopting the Galactic potential model \texttt{MWPotential2014}. We adopt (\texttt{r\_hi} $-$ \texttt{r\_lo})/2 as the error estimate of \texttt{r\_med}, corresponding to the 84th, 16th and 50th percentiles of photogeometric distance posterior respectively. Solar distance to the Galactic center and circular velocity at the Sun are set to 8 kpc and 220 \kms\ \citep{2012ApJ...759..131B}, consistent with the values adopted for the \texttt{MWPotential2014}. Finally, velocities $(U, V, W)$\footnote{U points towards the Galactic centre. V is along the direction of the Galactic rotation. W points at the North Galactic Pole.} are given relative to the Local Standard of Rest (LSR) using solar peculiar motion $(U_{\sun}, V_{\sun}, W_{\sun})$ = (11.1, 12.24, 7.25) \kms\ \citep{2010MNRAS.403.1829S}.

\begin{figure}
    \plotone{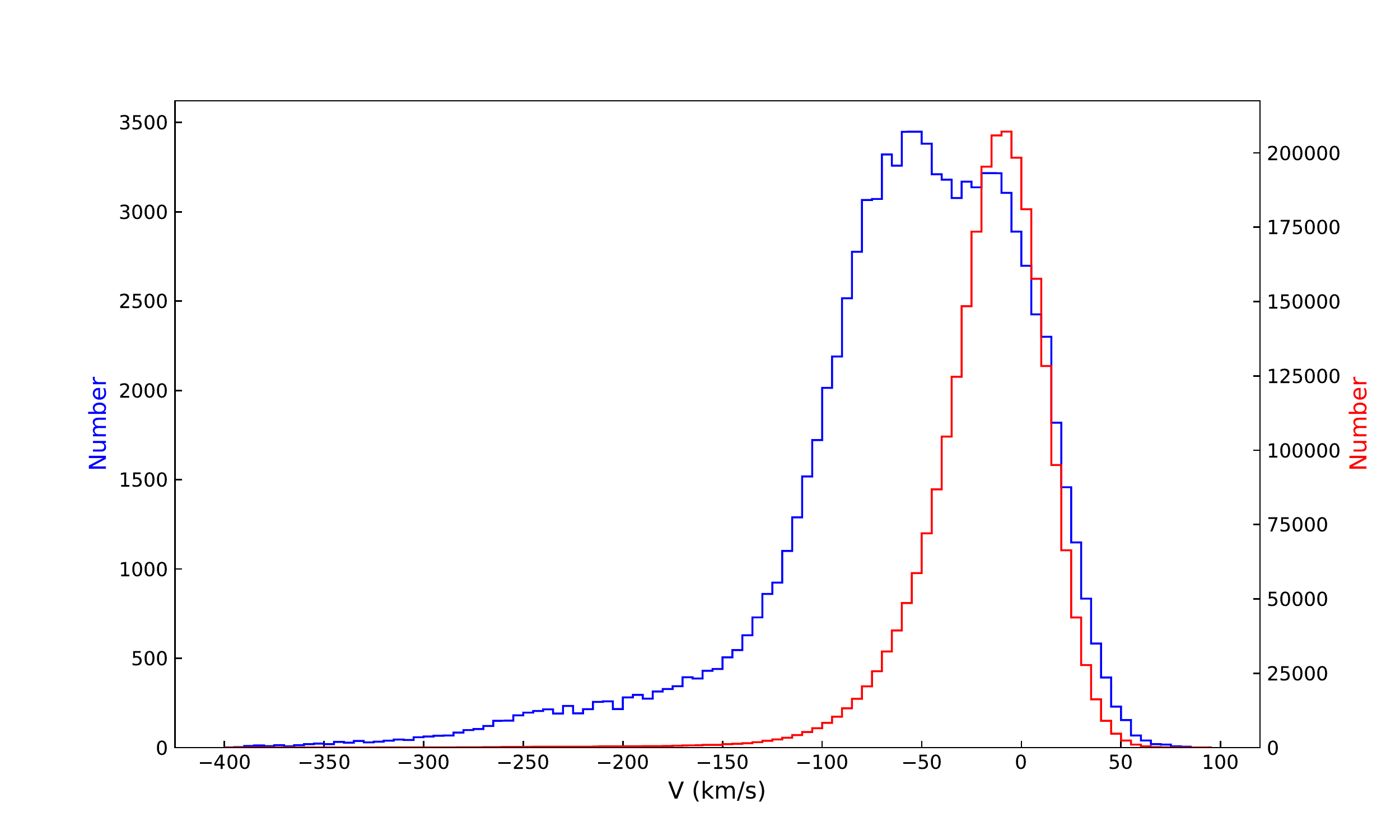}
    \caption{Histograms of $V$ velocity of LAMOST data. The red line represents the data with $\sigma_U,\sigma_V,\sigma_W$ \textless\ 10 \kms\ and $d$ \textless\ 2 kpc. The blue line denotes our final samples with $\sigma_U,\sigma_V,\sigma_W$ \textless\ 10 \kms, $d$ \textless\ 2 kpc and [Fe/H] \textless\ -0.7 dex.
    \label{fig:histsV}}
\end{figure}

For reduced data, we require that uncertainties of three velocity components $\sigma_U,\sigma_V,\sigma_W$ are \textless\ 10 \kms\ and distance $d$ is \textless\ 2 kpc. Considering that MGs of the thin disk have been well studied, we further select stars with [Fe/H] \textless\ -0.7 dex to mainly focus on the MGs of the thick disk\footnote{In this work, the terms ``thin disk" and ``thick disk" refer to the chemical thin and thick disks of the Milky Way (see Figure \ref{fig:chem}).}. It should be noted that the metallicity cut can not totally rule out the thin disk stars but the thick disk will become dominant after this cut. Our final samples consist of 91969 stars. Figure \ref{fig:histsV} shows the distributions of $V$ velocity of LAMOST data, in which the blue line denotes the selected samples and the red line represents the data without [Fe/H] limit. 

\section{Moving Groups Detection} \label{detection}
 
\subsection{Wavelet Transform}
The Wavelet transform (WT) technique can provide distinct signatures of substructures and has been widely used in the detection of MGs. If we denote the 2D distribution of data as $F(x,y)$, the WT coefficient at a certain point $(x',y')$ can be obtained by
\begin{equation}
    w(x',y') = \int^{\infty}_{-\infty}\int^{\infty}_{-\infty} \Psi(x,y)F(x,y) \,dxdy
    \label{equation1}
\end{equation} 
where $\Psi(x,y)$ is a Mexican hat function used as the mother wavelet, which is given by
\begin{equation}
    \Psi(x,y) = \left[ 2-\frac{(x-x')^2+(y-y')^2}{s^2} \right] {\rm exp} \left[ -\frac{(x-x')^2+(y-y')^2}{2s^2} \right]
\end{equation}
where $s$ is scale.

In this work, we perform WT on our samples in $V-\sqrt{U^2+2V^2}$ space. Exploring the MGs in $V-\sqrt{U^2+2V^2}$ was first proposed by \citet{2006A26A...449..533A} and successfully applied in \citet{2008ApJ...685..261K} and \citet{2014ApJ...787...31Z}. $V$ is proportional to vertical angular momentum $L_z$ and $\sqrt{U^2+2V^2}$ is related to orbital eccentricity $e$, in the Dekker's theory \citep{1976PhR....24..315D}.

\subsection{Construction of a Model Velocity Distribution}
An inevitable problem in MGs detection is Poisson noise. In \citet{2008ApJ...685..261K}, they created a ``smooth'' reference model velocity distribution that matched the overall velocities of the data. 250 Monte Carlo simulations were randomly drawn from the distribution, differences among which were due to Poisson noise. A feature detected in the data was compared to these simulations, in wavelet space, to see whether it was still significant. Here we employ the above method.

Considering the large number of our samples and asymmetries in velocities, especially in $V$ component, it is nearly impossible to design a Galactic model consisting of three Schwarzschild distributions that match the samples well. Instead, we run Gaussian mixture model, realized by extreme deconvolution (XD)\footnote{Available at \url{https://github.com/jobovy/extreme-deconvolution}} algorithm \citep{2011AnApS...5.1657B}, to construct the smooth velocity distribution. Bayesian Information Criterion \citep[integrated in \texttt{scikit-learn},][]{scikit-learn} is calculated to determine how many Gaussians are needed. The criterion decreases rapidly as the number of Gaussians is increased to 5, after which it stabilizes. Hence we run XD with 5 components and in this case, $U$, $V$ and $W$ are considered simultaneously. 

\begin{figure}[htb!]
    \plotone{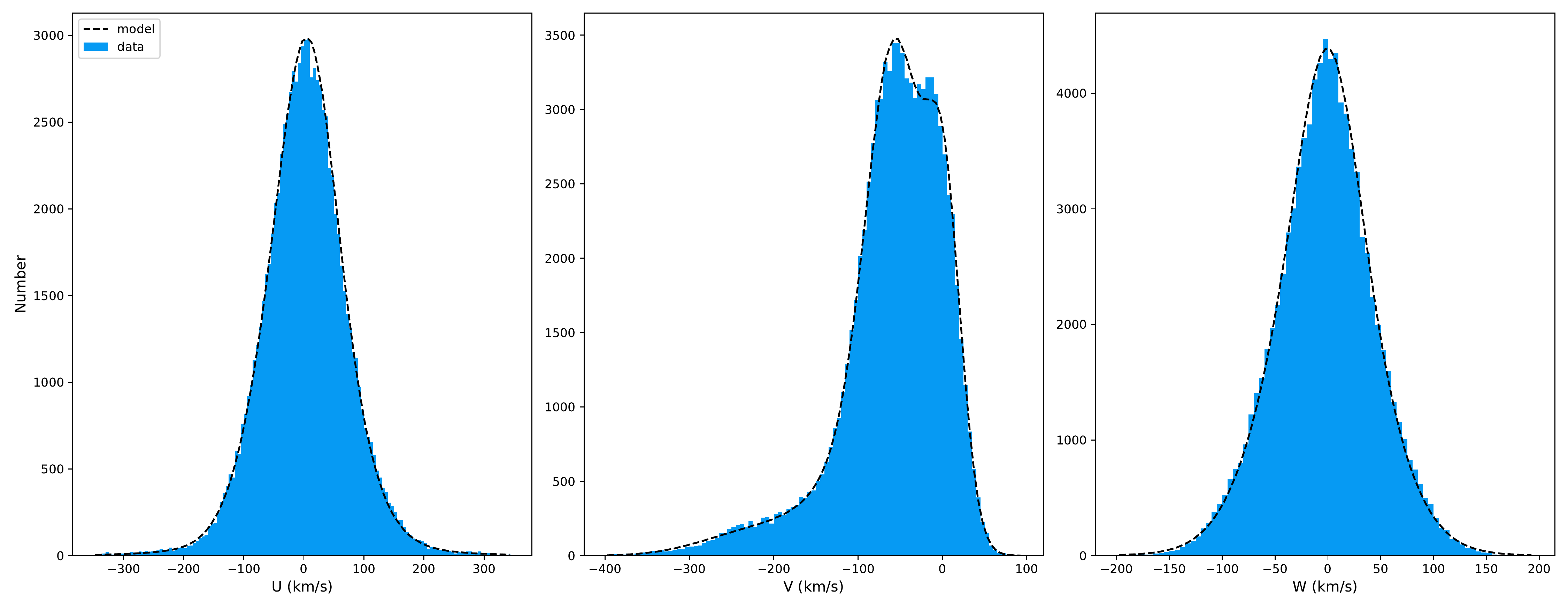}
    \caption{The model velocity distribution derived by XD and our samples.
    \label{fig:model}}
\end{figure}

The components derived by XD have mean ($U$, $V$) of (9.9, 2.4), (-9.1, -55.1), (-1.3, -55.6), (20.9, -96.9) and (2.0, -214.8) \kms, with corresponding weights of 0.23, 0.27, 0.29, 0.15 and 0.06 respectively. It can be seen that the Gaussian mixture is focusing on modelling the stars around the Sirius and Hercules groups. The model distribution is shown in Figure \ref{fig:model}, along with our samples as a comparison. Generally, the distribution fits the samples well, especially for the slope in the range between $V \simeq$ -200 \kms\ and -60 \kms, where the Galactic thick disk is roughly located. However, the model does not seem to be ideal below $V \simeq$ -200 \kms, which might be due to the fact that the stars here are quite sparsely distributed in velocity space. Hence we avoid speculating about this part.

\subsection{Detecting Moving Groups}

\begin{figure}[htb!]
    \plotone{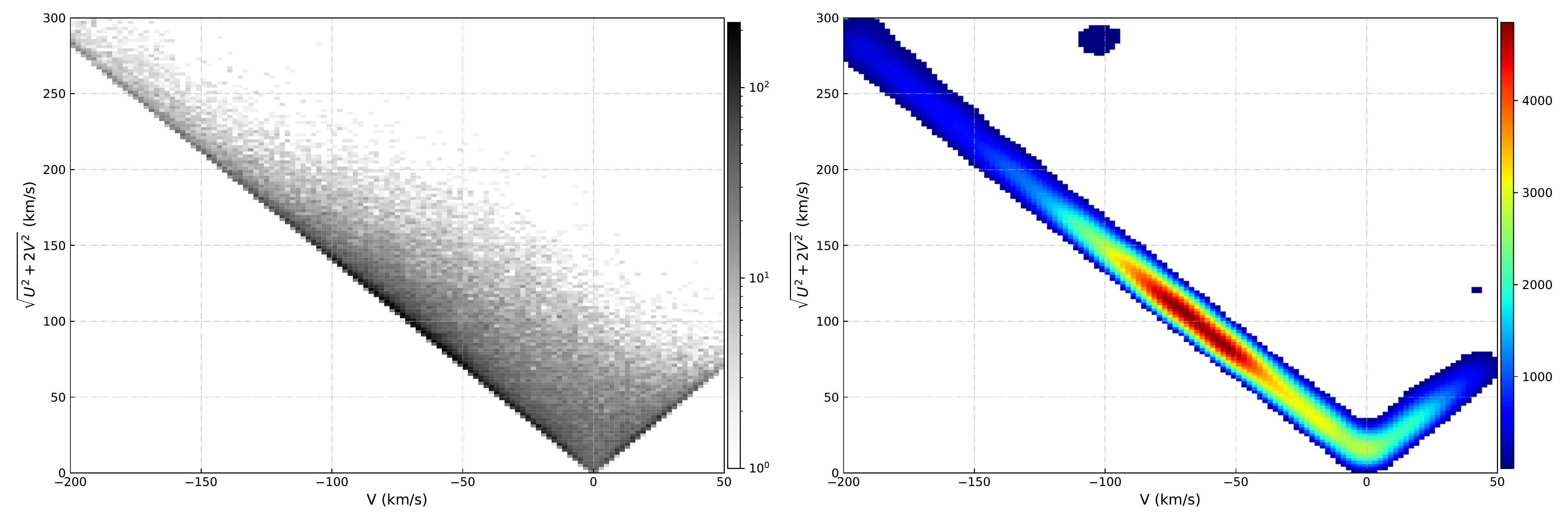}
    \caption{The left panel shows 2D histogram of the data in $V-\sqrt{U^2+2V^2}$ plane. The right panel shows the WT coefficients at scale $s$ = 10 \kms\ and only values $\geq 2$ are displayed.
    \label{fig:data_WT}}
\end{figure}

We bin the data in pixels of 2 \kms\ width in $V-\sqrt{U^2+2V^2}$ space and calculate the WT coefficient of each pixel $w_{data}$ through Equation \ref{equation1}. Since we are concerned about overdense regions, negative coefficients are set to zero. The left panel in Figure \ref{fig:data_WT} shows 2D histogram of the samples and the right one displays the corresponding WT coefficients at scale $s$ = 10 \kms.

\begin{figure}[htb!]
    \plottwo{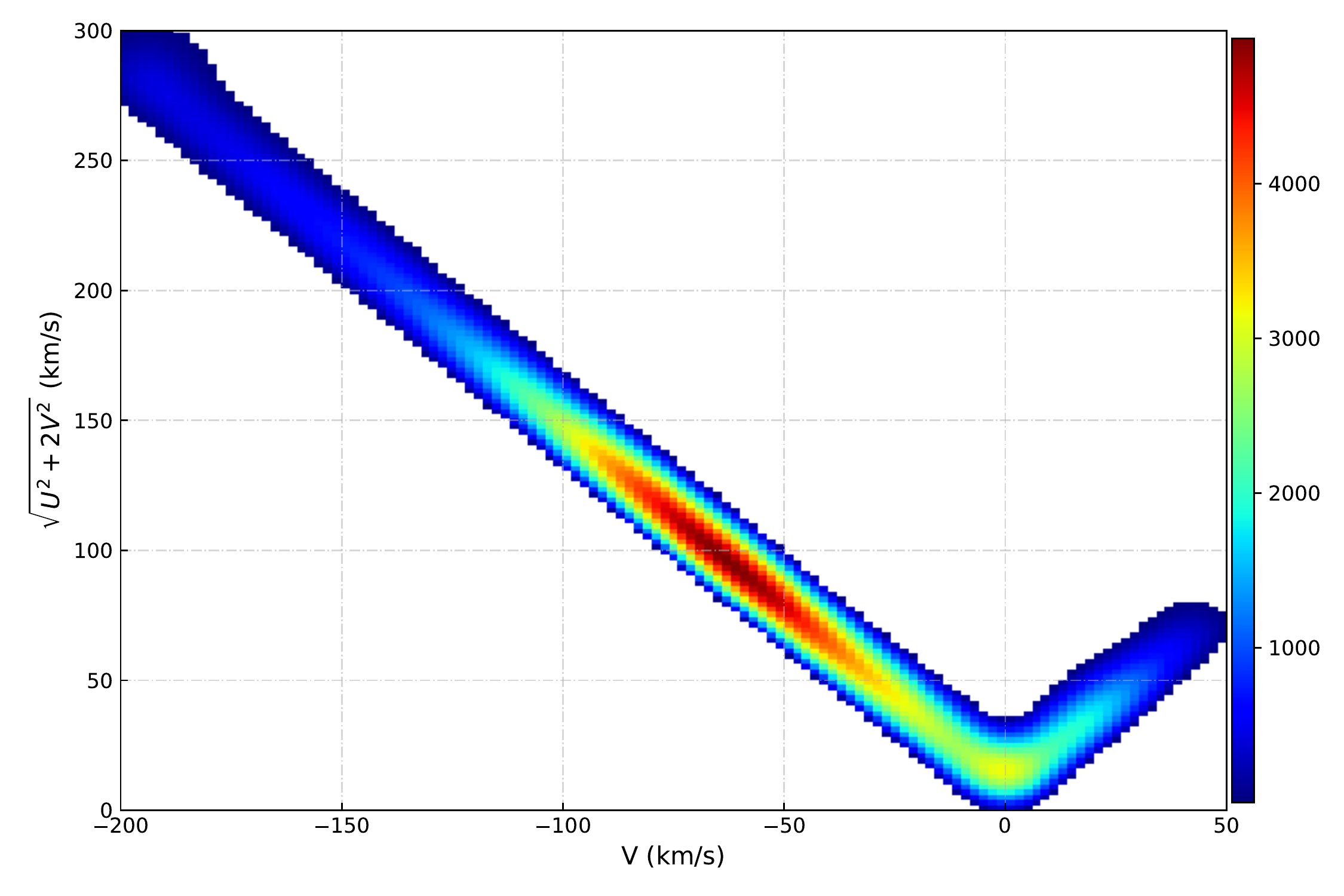}{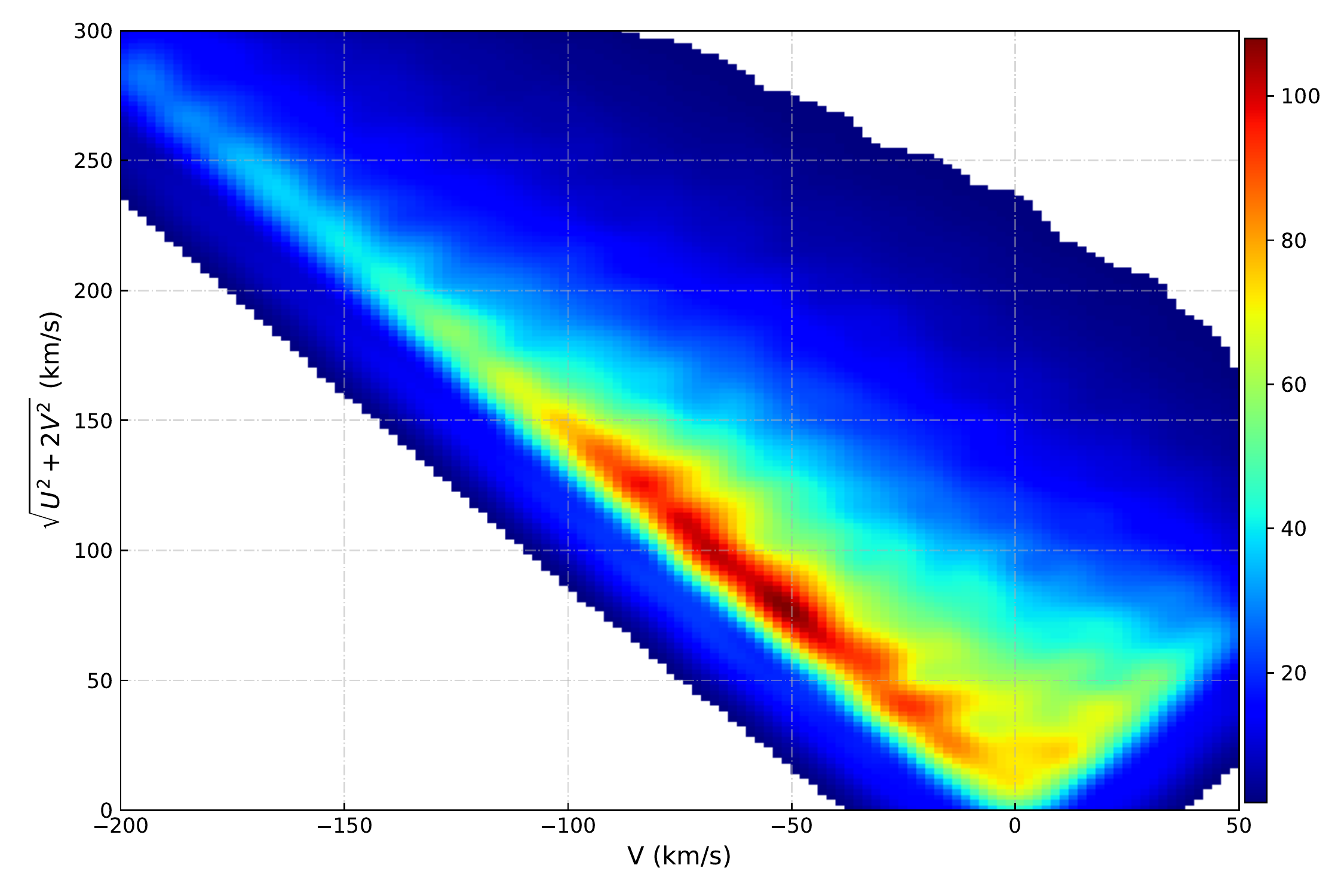}
    \caption{The left and right panels show the results of $\bar{w}_{MC}$ and $\sigma_{MC}$ at scale $s$ = 10 \kms, respectively. Only values $\geq 2$ are displayed.
    \label{fig:MC}}
\end{figure}

250 Monte Carlo (MC) simulations are generated by drawing, each time, 91969 mock stars from the model distribution built in the previous section. We proceed to apply the same WT on the simulations and calculate the mean $\bar{w}_{MC}$ and the standard deviation $\sigma_{MC}$ of the 250 MC WT coefficients in each pixel. $\bar{w}_{MC}$ is set to 0 whenever it is \textless\ 0 and $\sigma_{MC}$ is set to 1 whenever it is \textless\ 1, for the case that some pixels have no counts. Figure \ref{fig:MC} displays the WT of simulations at scale $s$ = 10 \kms. The left and right panels show the results of $\bar{w}_{MC}$ and $\sigma_{MC}$, respectively.

\begin{figure}[htb!]
    \plotone{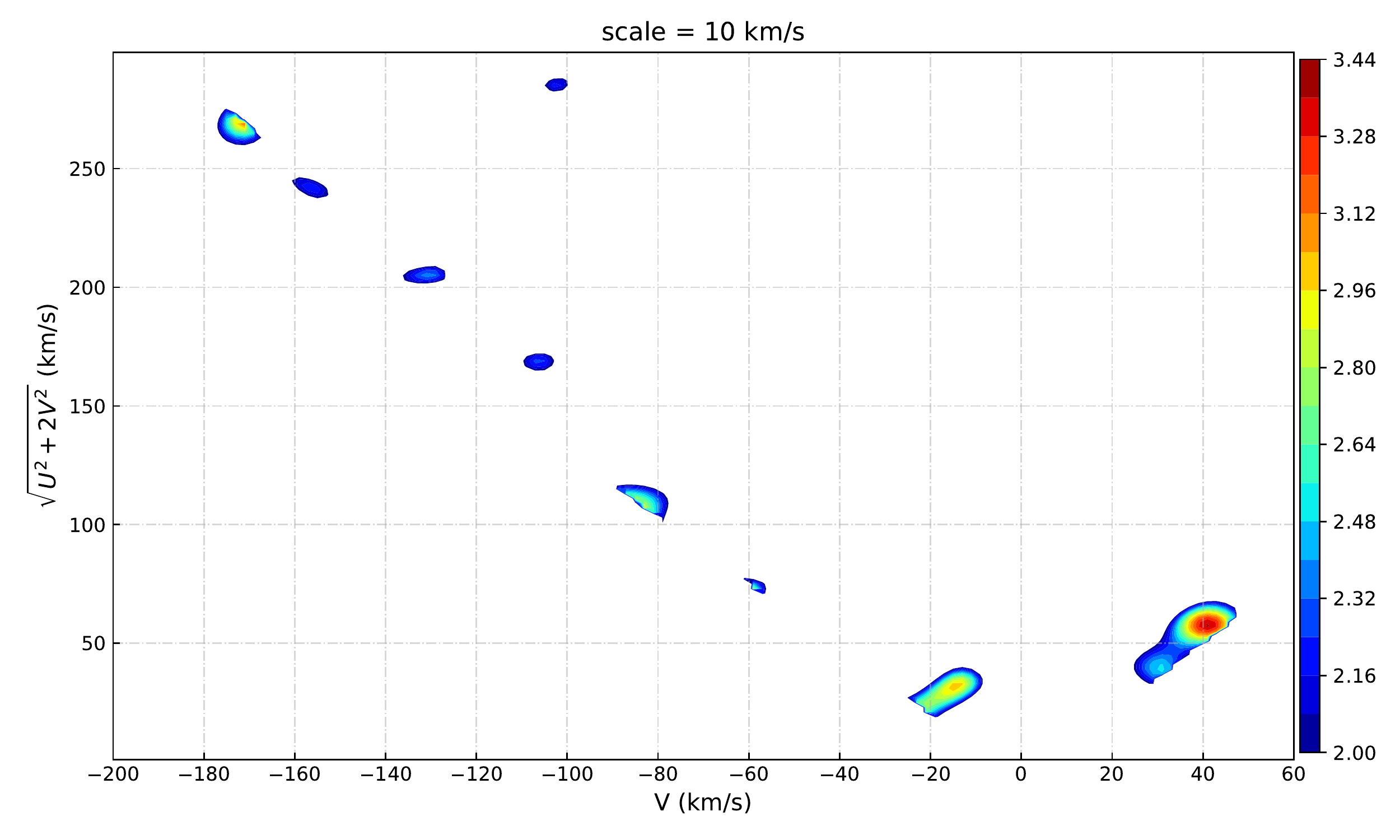}
    \caption{Contours of $\xi$ at scale $s$ = 10 \kms. Only values $\geq 2$ are displayed.
    \label{fig:s=10}}
\end{figure}

\begin{figure}[htb!]
    \plottwo{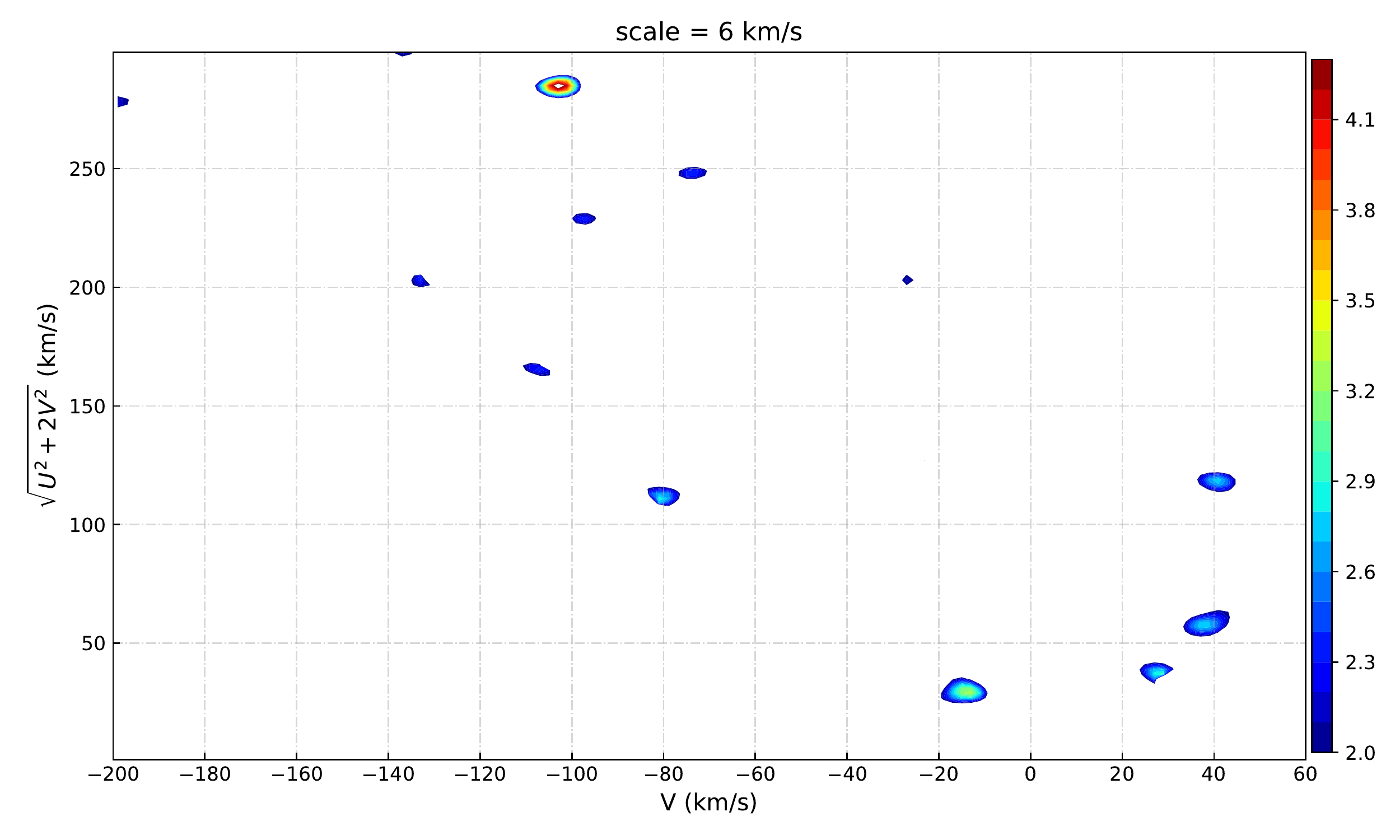}{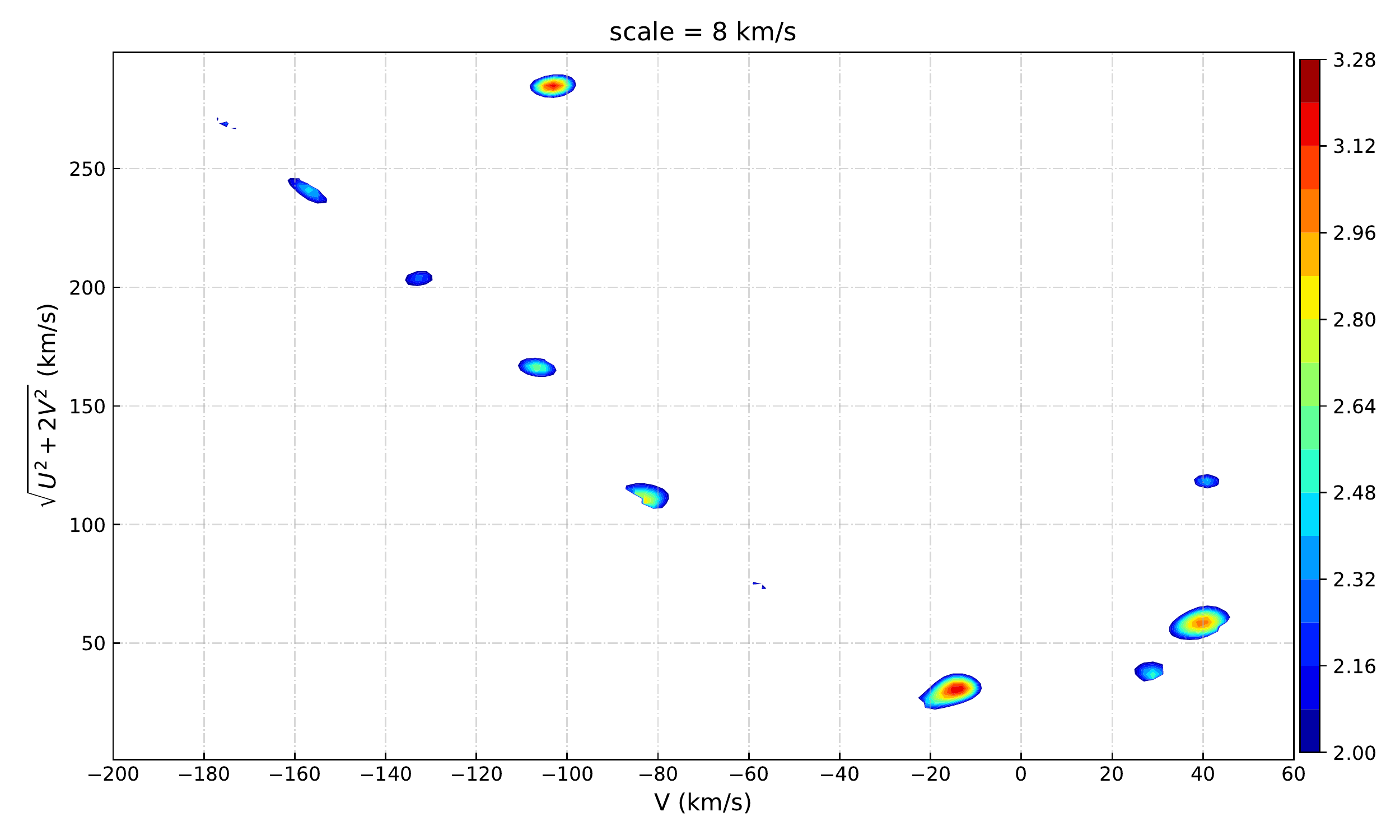}
    \plottwo{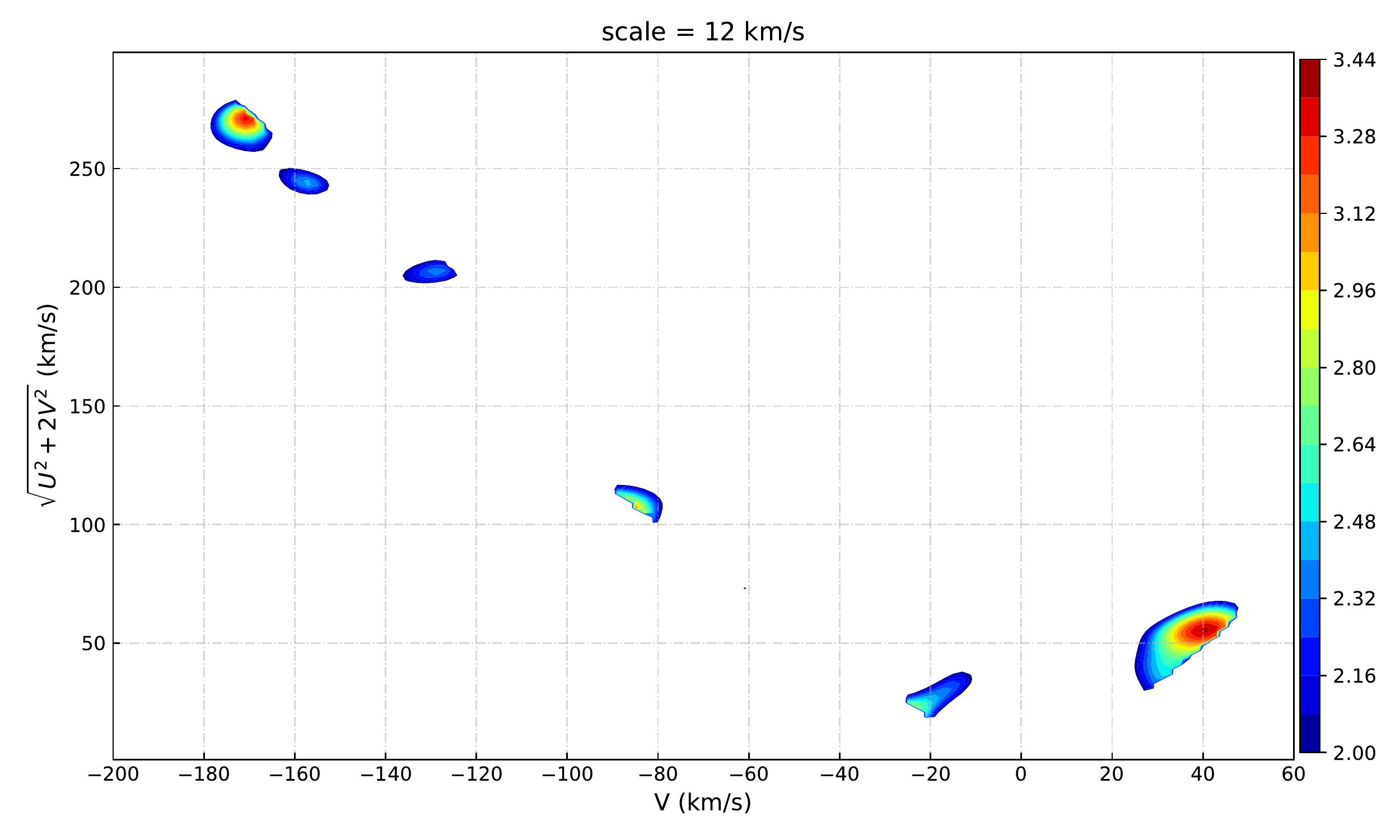}{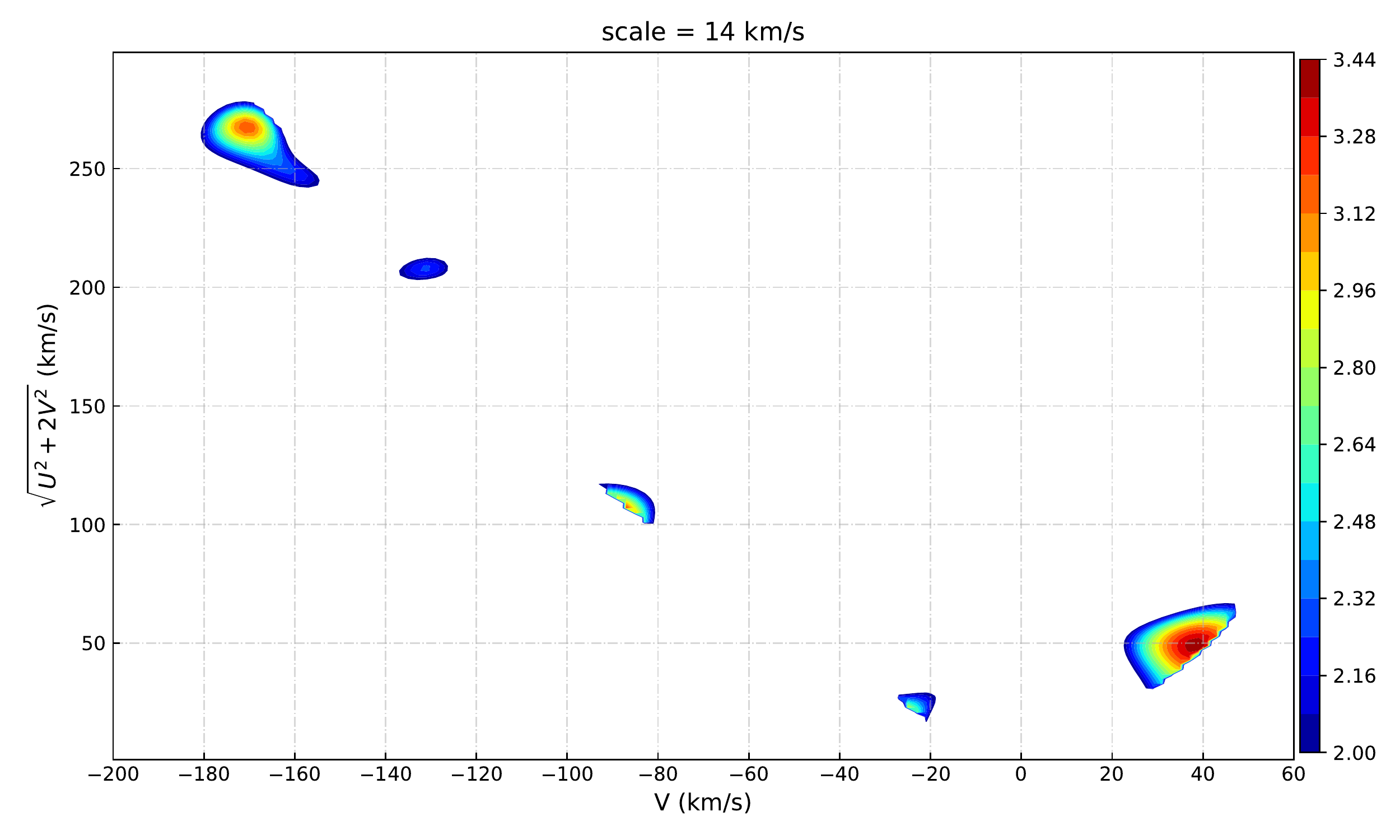}
    \caption{Contours of $\xi$ at scale $s$ = 6 \kms\ (upper left), $s$ = 8 \kms\ (upper right), $s$ = 12 \kms\ (lower left) and $s$ = 14 \kms\ (lower right). Only values $\geq 2$ are displayed.
    \label{fig:variouss}}
\end{figure}

The significance $\xi$ of a signal is defined as 
\begin{equation}
    \xi = \frac{w_{data} - \bar{w}_{MC}}{\sigma_{MC}}.
\end{equation}
We calculate $\xi$ at various scales and signals with $\xi \geq$ 2 are displayed in Figure \ref{fig:s=10} and \ref{fig:variouss}. We focus on the results at scale $s$ = 10 \kms\ since the structures of this scale are the most detectable. As is shown, several features stand out, some of which, however, should be treated with caution. Specifically, features in the range between $V \simeq$ -60 and 0 \kms\ are more likely caused by the smoothness of the model velocity distribution (see Figure \ref{fig:model}), which is also the reason why the Sirius and Hercules do not show up.

We match the remaining features with the MGs that have been studied in the literature. In the range of the thin disk, a MG with $V \simeq$ 40.7 \kms\ is detected, corresponding to ``A1'' found by \citet{2018A26A...619A..72R}. In the range of the thick disk, several overdensities arise in our LAMOST samples. The feature located at $V \simeq$ -82.4 \kms\ is described as AF06 \citep{2006A26A...449..533A}. At $V \simeq$ -105.5 \kms, there exists the Arcturus moving group \citep[e.g.,][]{2014A26A...562A..71B}. In addition, the feature at $V \simeq$ -156.0 \kms\ corresponds to KFR08 detected by \citet{2008ApJ...685..261K} and the one at $V \simeq$ -173.1 \kms\ corresponds to V3 in \citet{2014ApJ...787...31Z}. Note that the $V$ velocities of AF06, KFR08 and V3 are a little higher than the values in the literature\footnote{In the three works, $V_{\sun}$ = 5.2 \kms\ from \citet{1998MNRAS.298..387D} was used. KFR08 and V3 were centered at $V \simeq$ -160 and -180 \kms\ respectively. The center of AF06 was placed at $V \simeq$ -80 \kms\ but we think it should be $V \simeq$ -90 \kms\ because the mean and median of $V$ velocity of its potential members in their Table 2 are about -90 \kms.} because different solar peculiar velocities were used ($V_{\sun}$ = 12.24 \kms\ in this work). 

At $V \simeq$ -130.4 \kms, there is a new feature arising just as the other identified MGs. It is also detected at other scales shown in Figure \ref{fig:variouss}. To our knowledge it has not been confirmed yet. Given that we successfully detect the known MGs in the thick disk, we are able to claim that the feature located at $V \simeq$ -130.4 \kms\ should be a candidate for a new MG. 

Now we summarize the known MGs as follows. In \citet{2018A26A...619A..72R}, two new arches are detected: one is ``A1'' with $V \simeq$ 38 \kms\ and the other is ``A2'' with $V \simeq$ 15 \kms. The Sirius has $V \simeq$ 0 \kms. The Pleiades/Hyades stream is located between $V \simeq$ -10 and -20 \kms. The Hercules has $V$ between -40 $\sim$ -50 \kms. HR1614 is at $V \simeq$ -65 \kms. Together with AF06, the Arcturus, the new feature in this work, KFR08 and V3 as described above, we conclude that MGs in the Galactic disk are spaced by approximately 15 $\sim$ 25 \kms\ along $V$ velocity, taking into account velocity uncertainties and sizes of the structures. 

\section{Chemistry and Age} \label{chem_age}

\begin{figure}[htb!]
	\plotone{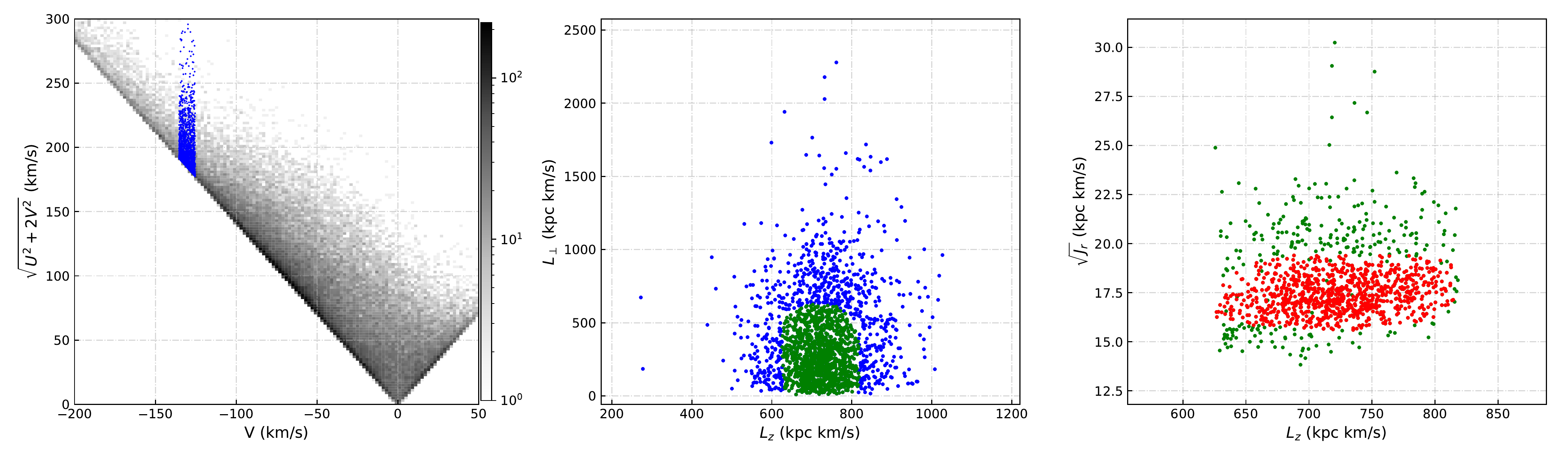}
	\caption{An example of member stars selection for the new detected feature. Blue dots represent the stars selected based on $V$ velocity. Green dots indicate the stars selected in $L_z-L_{\bot}$ space. The chosen member stars are marked with red dots.
		\label{fig:selection}}
\end{figure}

We investigate the detected MGs through distributions of chemistry and age, based on APOGEE DR16 and isochrone ages from \citet{2018MNRAS.481.4093S}. For the data, $\sigma_U,\sigma_V,\sigma_W$ \textless\ 10 \kms\ and $d$ \textless\ 2 kpc are required. Then the candidate member stars of each MG are selected progressively in three planes defined by combinations of velocity, angular momentum and action components: $V-\sqrt{U^2+2V^2}$ plane, $L_z-L_{\bot}$ plane \citep{1999Natur.402...53H} and $L_z-\sqrt{J_r}$ plane \citep[][where $J_r$ is radial action]{2019MNRAS.484.3291T}. This will give us stronger criteria on the selection of member stars. Figure \ref{fig:selection} illustrates an example of this selection for the new detected feature. Stars within 5 \kms\ around $V$ = -130.4 \kms\ are selected (blue dots). They are explored in $L_z-L_{\bot}$ space and stars within the most concentrated region are picked out (green dots). These stars are further plotted in $L_z-\sqrt{J_r}$ plane and ones in the densest area are chosen as the members (red dots). The aim of this procedure is to select the stars concentrated in kinematics and dynamics, although it is somewhat subjective since the level of concentration is judged by eye.

\begin{figure}[htb!]
    \plotone{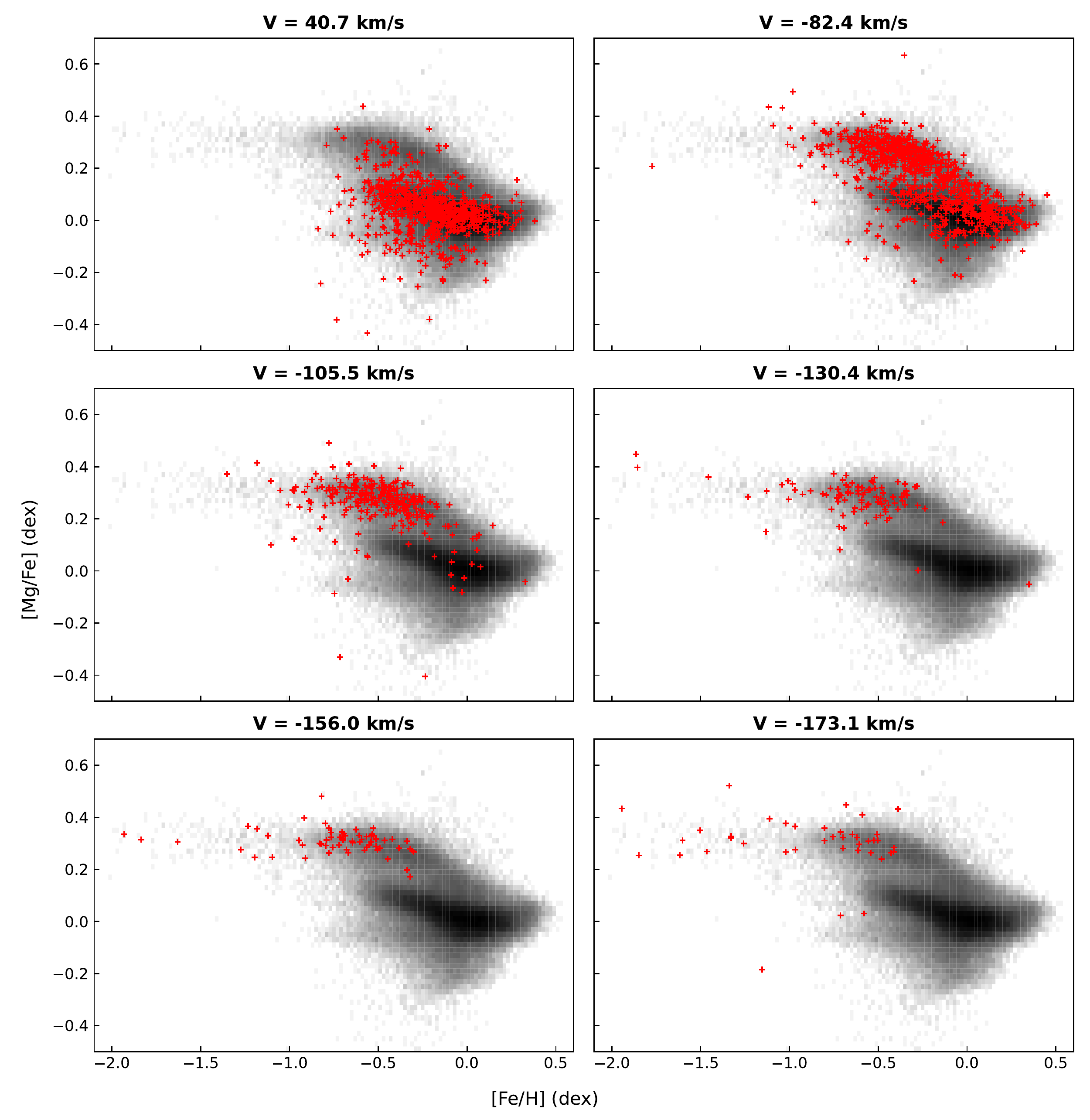}
    \caption{[Fe/H]$-$[Mg/Fe] trends of MGs (red cross) and the Milky Way (gray background) using APOGEE data. Corresponding $V$ velocity is marked at the top of each panel.
    \label{fig:chem}}
\end{figure}

Figure \ref{fig:chem} shows the distributions of MGs members in [Fe/H]$-$[Mg/Fe] plane. Apparently, ``A1'' belongs to the thin disk alone. AF06 is a mixture of the thin and thick components. Arcturus, together with the new feature, KFR08 and V3, is related to the thick disk and even halo. In addition to these MGs, we also know that the Sirius and the Pleiades/Hyades are the ones in the thin disk \citep[e.g.,][]{2014IAUS..298...77R,2007A26A...461..957F}, and the Hercules as well as HR1614 is a mixture of thin and thick disk stars \citep[e.g.,][]{2007ApJ...655L..89B,2020A26A...638A.154K}. 
It can be seen that MGs have gone through a coherent transitional process from only containing the thin disk stars, to being a mixture of the thin and thick disks, and then to containing the thick disk and even halo stars. This trend is reasonable because the different Galactic components have different rotational offsets from the LSR \citep[e.g.,][]{2003A26A...410..527B}. The thin disk stars have higher $V$ while the thick disk stars have lower $V$. The Hercules, HR1614 and AF06 are located in the range of a mixture of the thin and thick disks. What's more, there is no sign indicating that any of these MGs follows a distinct chemical sequence from the background Milky Way stars. So we can infer that MGs mentioned here should not be treated as remnants of accreted galaxies.

\begin{figure}[htb!]
    \plotone{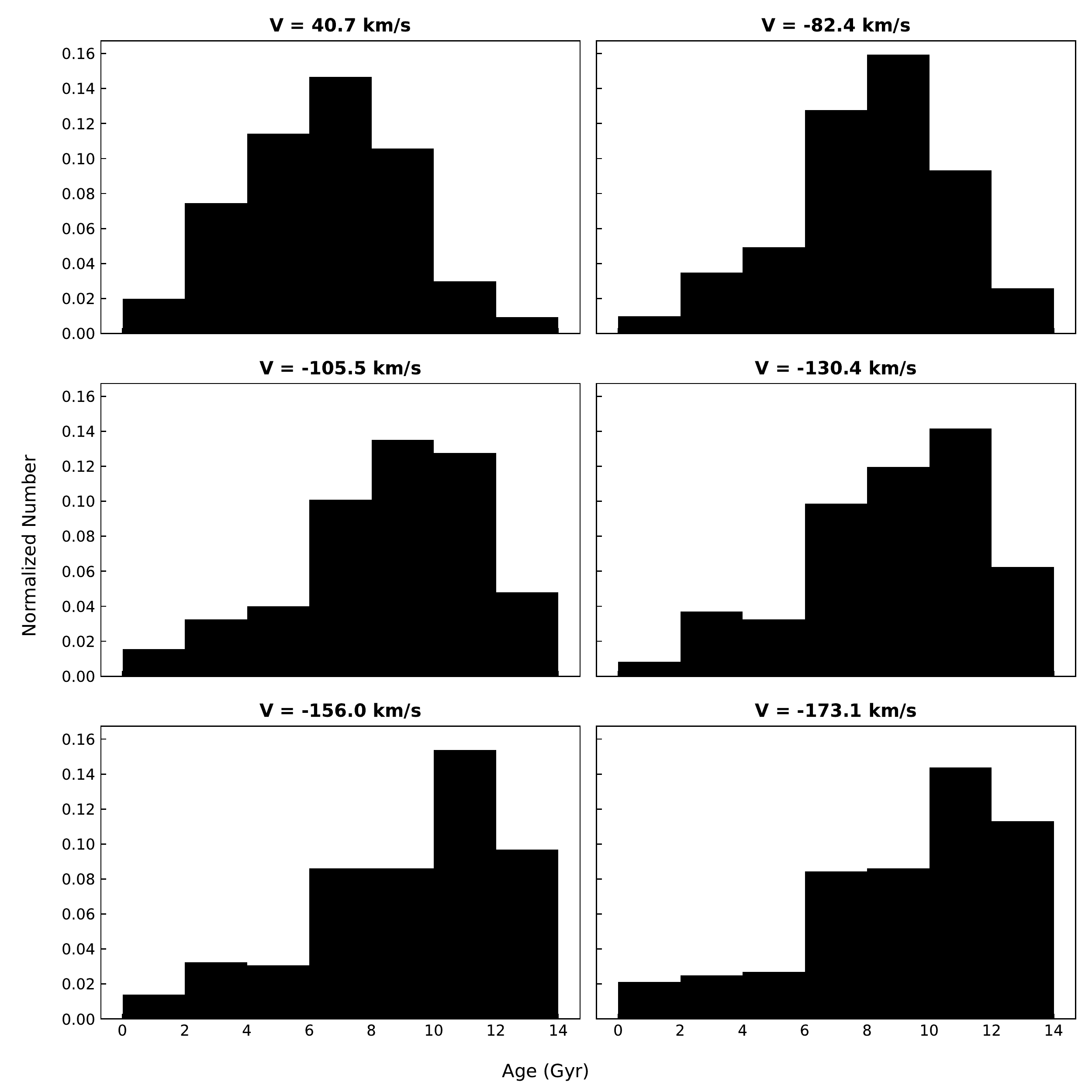}
    \caption{Histograms of age for MGs. The isochrone ages come from \citet{2018MNRAS.481.4093S}. Each histogram is normalized which means that the enclosed area is equal to 1.
    \label{fig:age}}
\end{figure}

The histograms of age are presented in Figure \ref{fig:age}, for which the errors are less than 2 Gyr. As is shown, MGs with lower $V$ tend to be dominated by older stars, which is caused by different concentrations of age of stars populated in different Galactic components. It is worth noting that very young stars (\textless\ 2 Gyr) are contained in the MGs. \citet{2009MNRAS.396L..56M} attributed the origin of MGs with low angular momenta to the dynamical perturbation by a merger event occurred $\sim$ 1.9 Gyr ago, which required that these MGs should be, on average, older than the time. This hypothesis is disfavoured considering the age distributions here. However, we can not totally exclude the mechanism of \citet{2009MNRAS.396L..56M} based only on an age argument because younger stars might be born with similar kinematic and dynamic properties to the MGs.

\section{Conclusion and diskussion} \label{summary}

We detect MGs in the solar neighborhood with a sample of 91969 nearby stars constructed from LAMOST DR7. The origins of MGs are analysed through chemistry and age.

One candidate for a new MG is detected using the wavelet technique and Monte Carlo simulations. The new feature is centered at $V \simeq$ -130 \kms. Together with other known substructures, we conclude that MGs in the Galactic disk are spaced by approximately 15 $\sim$ 25 \kms\ along $V$ velocity.

The wide spreads of chemical abundances and ages within MGs can rule out their dissolved open cluster origin. No existences of distinct [Fe/H]$-$[$\alpha$/Fe] trends from the background Milky Way stars exclude the remnants of accreted galaxies. The hypothesis of perturbation induced by a past merger is disfavoured given that very young stars are contained in MGs. It is the resonant mechanism that does not contradict the results here. Therefore, we attribute the origin of MGs to the continuous resonances caused by the Galactic bar or spiral arms of the Milky Way.

In chemistry, the compositions of MGs change from the thin disk stars, to mixtures of the thin and thick disks, and then to the thick disk and even halo stars. In terms of age, MGs with lower $V$ tend to be older than those with higher $V$. This seems to be a coherent transitional process, implying that they might be linked together rather than treated separately. 

AF06 used to be considered as a MG in the thick disk. With [Fe/H]$-$[$\alpha$/Fe] information here, it is clear that AF06 contains both the thin and thick disk stars. 

Why the MGs are spaced by 15 $\sim$ 25 \kms\ along $V$? There should be a mechanism that will trap stars in velocity space as long as their motions coincide with some certain condition. Thus, answering what the mechanism is will be the key to uncover the mystery of MGs. In addition, some clump substructures with low angular momenta were reproduced in $U - V$ space in \citet{2009ApJ...700L..78A} considering the bar and/or spiral arms together with a hotter disk as the initial condition ((f) and (i) panel in their Figure 1). Will it be a good scenario if we connect ``hotter disk" to the Galactic thick disk that was once heated by the Gaia-Enceladus-Sausage \citep{2018MNRAS.478..611B,2018Natur.563...85H}? Modeling how the MGs with low angular momenta are generated should also play a vital role.

\begin{acknowledgments}
We thank the anonymous referee, whose comments greatly improved this publication. We acknowledge the support from the 2m Chinese Space Station Telescope project: CMS-CSST-2021-B05. This research made use of TOPCAT, an interactive graphical viewer and editor for tabular data \citep{2005ASPC..347...29T}. This study is supported by the National Natural Science Foundation of China under grant No. 11988101, 11973048, 11927804, 11890694 and National Key R\&D Program of China No. 2019YFA0405502. 

Guoshoujing Telescope (the Large Sky Area Multi-Object Fiber Spectroscopic Telescope LAMOST) is a National Major Scientific Project built by the Chinese Academy of Sciences. Funding for the project has been provided by the National Development and Reform Commission. LAMOST is operated and managed by the National Astronomical Observatories, Chinese Academy of Sciences.

This work presents results from the European Space Agency (ESA) space mission Gaia. Gaia data are being processed by the Gaia Data Processing and Analysis Consortium (DPAC). Funding for the DPAC is provided by national institutions, in particular the institutions participating in the Gaia MultiLateral Agreement (MLA). The Gaia mission website is \url{https://www.cosmos.esa.int/gaia}. The Gaia archive website is \url{https://archives.esac.esa.int/gaia}.

Funding for the Sloan Digital Sky Survey IV has been provided by the Alfred P. Sloan Foundation, the U.S. Department of Energy Office of Science, and the Participating Institutions. 

SDSS-IV acknowledges support and resources from the Center for High Performance Computing  at the University of Utah. The SDSS website is \url{www.sdss.org}.

SDSS-IV is managed by the Astrophysical Research Consortium for the Participating Institutions of the SDSS Collaboration including the Brazilian Participation Group, the Carnegie Institution for Science, Carnegie Mellon University, Center for Astrophysics | Harvard \& Smithsonian, the Chilean Participation Group, the French Participation Group, Instituto de Astrof\'isica de Canarias, The Johns Hopkins University, Kavli Institute for the Physics and Mathematics of the Universe (IPMU) / University of Tokyo, the Korean Participation Group, Lawrence Berkeley National Laboratory, Leibniz Institut f\"ur Astrophysik Potsdam (AIP),  Max-Planck-Institut f\"ur Astronomie (MPIA Heidelberg), Max-Planck-Institut f\"ur Astrophysik (MPA Garching), Max-Planck-Institut f\"ur Extraterrestrische Physik (MPE), National Astronomical Observatories of China, New Mexico State University, New York University, University of Notre Dame, Observat\'ario Nacional / MCTI, The Ohio State University, Pennsylvania State University, Shanghai Astronomical Observatory, United Kingdom Participation Group, Universidad Nacional Aut\'onoma de M\'exico, University of Arizona, University of Colorado Boulder, University of Oxford, University of Portsmouth, University of Utah, University of Virginia, University of Washington, University of Wisconsin, Vanderbilt University, and Yale University.
\end{acknowledgments}

%





\bibliography{sample631}{}
\bibliographystyle{aasjournal}



\end{document}